\documentclass{iau}

\usepackage{amsmath}
\usepackage{graphicx}
\usepackage{multirow}
\usepackage{csquotes}

\begin{document}

\lefttitle{M. A. Garrett}
\righttitle{SETI's blind spot: Technological Acceleration \& fleeting technosignatures}

\jnlPage{1}{4}
\jnlDoiYr{2026}
\doival{10.1017/S1743921326XXXXX}

\aopheadtitle{Proceedings IAU Symposium No. 404}
\editors{J. Haqq-Misra \& R. Kopparapu, eds.}

\title{SETI's blind spot: Technological Acceleration \& fleeting technosignatures}

\author{Michael A. Garrett}
\affiliation{Jodrell Bank Centre for Astrophysics, Department of Physics \& Astronomy, Oxford Road,
Alan Turing Building, University of Manchester, M13 9PL, UK \\
email: {\tt michael.garrett@manchester.ac.uk}}

\begin{abstract}
The search for extraterrestrial intelligence (SETI) has traditionally framed the detection challenge with a focus on the parameter $L$ in the Drake equation - the communicative lifetime
of a civilisation.  I argue that the more pertinent quantity is $\tau_d$, the duration
during which a civilisation produces technosignatures that are \emph{actually detectable by us, now}.
Modelling technological progress as an exponential process,
we show that $\tau_d = \alpha^{-1}\ln(K_{\rm max}/K_{\rm min})$, where $\alpha$ is the
rate of technological acceleration and $[K_{\rm min},K_{\rm max}]$ brackets the technology
levels accessible to our instruments.  As $\alpha$ increases, $\tau_d$ can shrink to mere decades, dramatically narrowing the
window in which civilisations overlap technically. This
``technology mismatch'' has implications for future search strategies, emphasising
broadband and technology-agnostic approaches, as well as anomaly detection across
multi-wavelength/messenger survey data.
\end{abstract}

\begin{keywords}
extraterrestrial intelligence, technosignatures, SETI, technological acceleration,
Great Silence, radio leakage
\end{keywords}

\maketitle

\section{Introduction}
Drake's original formulation of the number of detectable civilisations in the Galaxy
(\cite{Drake1961}) has often been reduced to $N \sim L$, where $L$ is the communicative
lifetime of a civilisation.  This framing implicitly assumes that a civilisation remains
detectable throughout its communicative phase, i.e.\ that its technosignatures always
overlap in character with our search capabilities.  In 1973, Sagan recognised a
fundamental difficulty: civilisations substantially more advanced than our own ``will have
technologies and laws of nature currently inaccessible to us''
\cite{Sagan1973}.  He introduced the concept of a \emph{detection window}, $\tau_d$,
suggesting that the more relevant Drake term is not $L$ but the shorter interval during
which a civilisation's signatures are actually detectable --- leading to $N \sim \tau_d$
rather than $N \sim L$.

This technology mismatch perhaps becomes clearer when we consider the contemporary
trajectory of our own civilisation.  Radio communications offer a telling example.
Between roughly 1930 and 1990, Earth's electromagnetic signature was dominated by a
small number of high-power, low-gain, narrowband, analogue broadcast transmitters -
precisely the kind of signal that traditional and ongoing SETI efforts have always targeted.  By contrast,
the period from 1990 onwards has been characterised by exactly the oppossite - a large number of low-power,
high-gain, broadband, digital devices \cite{Sullivan1978,Saide2023}.  Most current SETI
efforts are still focused on narrowband signals in the style of the earlier era, and are
therefore already partially mismatched to our own present-day technology. This limits us to the detection of what be described as "primitive", deliberate, narrowband beacons.  

\section{The Detection Window Model}

\subsection{Exponential technological growth}

Let the technological level of a civilisation be described by a continuous variable $K$
that increases with time $t$.  We model progress as an exponential process
\cite{Kurzweil2005}:
\begin{equation}
  K(t) = K_0\,e^{\alpha t},
  \label{eq:K}
\end{equation}
where $K_0$ is the initial level at $t=0$ and $\alpha$ (yr$^{-1}$) is the rate of
technological acceleration.  For $\alpha \ll 1$, equation~(\ref{eq:K}) reduces to
linear growth.  Historically, individual technologies follow logistic (S-curve)
trajectories, but the aggregate capability of a civilisation can be approximated as a
superposition of many overlapping S-curves.  Historically, individual technologies often follow logistic (S-curve) trajectories, with periods of rapid growth followed by saturation \cite{Foster1986}, but the aggregate capability of a civilisation can be approximated as a superposition of many overlapping S-curves.

Artificial intelligence represents a
qualitative departure: unlike prior domain-specific technologies, AI acts as a
meta-technology that accelerates innovation across all other domains, including itself,
potentially driving $\alpha$ to values far larger than any previously observed.

\subsection{Duration of the detection window}

Our ability to detect the technosignatures of another civilisation is limited to a
specific range of technology levels, $[K_{\rm min}, K_{\rm max}]$.  Signatures from
technologies below $K_{\rm min}$ are too faint or primitive to be observed at
interstellar distances; those above $K_{\rm max}$ rely on physical principles or media
currently unknown to us, rendering them invisible to our instruments.

From equation~(\ref{eq:K}), the civilisation reaches $K_{\rm min}$ and $K_{\rm max}$
at times $t_1 = \alpha^{-1}\ln(K_{\rm min}/K_0)$ and
$t_2 = \alpha^{-1}\ln(K_{\rm max}/K_0)$ respectively.  The duration of the detectable
window is therefore
\begin{equation}
  \tau_d \;=\; t_2 - t_1 \;=\; \frac{1}{\alpha}\ln\!\left(\frac{K_{\rm max}}{K_{\rm min}}\right).
  \label{eq:tau}
\end{equation}

Equation~(\ref{eq:tau}) shows that $\tau_d$ is \emph{inversely proportional} to the
acceleration rate $\alpha$, but only \emph{logarithmically} sensitive to the technology
range ratio $K_{\rm max}/K_{\rm min}$.  As $\alpha$ grows, the detection window
narrows; the precise value of the dynamic range makes little quantitative difference
(see Section~\ref{sec:estimates}). We note that our analysis assumes a constant technological acceleration rate $\alpha$. More generally, if $\alpha$ varies with time, the detection window is determined by the integrated growth history. Equation~(\ref{eq:tau})  should therefore be regarded as an effective first-order approximation based on the mean acceleration rate over the detectable interval.

\section{Estimating Plausible Values of $\tau_d$}
\label{sec:estimates}

The key parameter is $\alpha = \ln 2 / T_2$, where $T_2$ is the technology doubling
time.  Drawing on historical trends \cite{Garrett2026a}, we define \emph{Slow growth:} ($T_2 \sim 100$ yr, pre-industrial), $\alpha \approx 0.007$ yr$^{-1}$; 
 \emph{Moderate growth} ($T_2 \sim 25$ yr, late 20th century), $\alpha \approx 0.03$ yr$^{-1}$;
   \emph{Rapid growth} ($T_2 \sim 5$ yr, early AI era), $\alpha \approx 0.14$ yr$^{-1}$; and
  \emph{Post-biological / ASI-driven growth} ($T_2 < 1$ yr): $\alpha > 1$ yr$^{-1}$.

A ratio $K_{\rm max}/K_{\rm min} \sim 10^6$ is a conservative estimate for the
dynamic range of a single technological paradigm before it becomes obsolete, giving
$\ln(K_{\rm max}/K_{\rm min}) \approx 14$.  Under these assumptions:
$\tau_d \approx 2000$ yr (slow), $\approx 500$ yr (moderate), $\approx 100$ yr (rapid),
and $\lesssim 20$ yr (post-biological with $\alpha \geq 1$ yr$^{-1}$). Choosing different values of $K_{\rm max}/K_{\rm min}$ makes little difference to these figures. 



A selection effect follows directly from equation~(\ref{eq:tau}): SETI surveys are
inherently biased toward detecting civilisations with low $\alpha$, since their longer
windows $\tau_d$ make them statistically easier to find.  Rapidly advancing
civilisations are the
hardest to catch.

\section{Caveats}

The model is deliberately simple, and several factors could extend $\tau_d$ beyond
the values derived above. These include: 

\emph{Persistence and legacy signals.}  Legacy beacons designed to operate autonomously
for millennia, and engineered artefacts such as megastructures, may outlive the
civilisations that built them \cite{Cirvovic2006}.  Infrastructure inertia can also
slow the replacement of older, more detectable technologies \cite{Hughes1983}.

\emph{Physical limits.}  Thermodynamic constraints ultimately cap energy consumption
and computational growth \cite{Lloyd2000}.  Latency across large distributed networks
imposes additional practical bounds on the coherence of any distributed
super-intelligence.

\emph{Behavioural factors.}  Civilisations may deliberately plateau their technological
development for cultural, political, or environmental reasons.  Alternatively, some advanced
societies might enhance rather than abandon powerful beacons, maintaining or widening
$\tau_d$ for intentional signals even as their internal braodband communications become
undetectable.

These moderating factors notwithstanding, the central result remains robust: rapid
technological acceleration compresses $\tau_d$ significantly.  For post-biological
civilisations with $\alpha \geq 1$ yr$^{-1}$, detectable windows become
cosmologically negligible, reframing the Great Silence not as evidence of absence but
as evidence of extreme technological disparity.

\section{Implications for SETI Strategies}

If $\tau_d$ is indeed short for most advanced civilisations, the traditional focus on
narrowband radio signals is doubly constrained: such signals are already a product of a
transitional technological phase on Earth, and the equivalent phase elsewhere may be
vanishingly brief.  Several strategic conclusions follow.

\emph{Broadband radio searches.}  In addition to narrowband SETI, searches should
expand to cover the full accessible frequency domain (MHz--THz), consistent with the
leakage profile of a more mature technological civilisation.  Simulations show that
Earth's aggregate mobile and radar emission would constitute a detectable broadband
source to a sufficiently advanced receiver \cite{Saide2023}, motivating analogous
searches at extrasolar distances. Most recently, arguments have been made to begin deep, wide-field broadband radio surveys of the sky with SKA and ngVLA pathfinders to detect BRaTs (Broadband Radio Technosignatures), using multi-wavelength data and AI to eliminate confounding sources \citep{Garrett2026b}. 

\emph{Technology-agnostic, persistent signatures.}  More effort should be directed
toward signatures whose lifetimes are $\gg \tau_d$: waste heat from megastructures,
Dyson sphere candidates, anomalous stellar dimming, and other large-scale
astro-engineering manifestations.  Such signatures are rooted in energy conservation
and are therefore detectable regardless of the specific technology producing them.
Wide-field surveys with facilities such as the Vera C.~Rubin Observatory are well placed to search systematically for such rare anomalies.

\emph{AI-driven anomaly detection.}  Future SETI is, in essence, an anomaly-detection
problem across high-dimensional, multi-wavelength and multi-messenger survey data.
Machine learning already offers powerful tools for distinguishing non-natural outliers
in electromagnetic datasets.  As AI capabilities grow, the ability to construct
accurate models of the natural universe and to flag departures from expectation will
expand enormously, potentially identifying technosignatures too subtle or
non-anthropocentric for traditional methodologies \citep{Garrett2026b}. 

\section{Conclusions}

We have presented a simple model in which the duration of detectability of a
civilisation's technosignatures is
$\tau_d = \alpha^{-1}\ln(K_{\rm max}/K_{\rm min})$.  The key results are:

\begin{enumerate}
  \item $\tau_d$ is inversely proportional to the technological acceleration rate
        $\alpha$ and only weakly sensitive to the dynamic range of detectable
        technology.
  \item For rapid AI-driven growth ($\alpha \approx 0.14$ yr$^{-1}$, $T_2 \sim 5$ yr),
        $\tau_d \approx 100$ yr.  For post-biological civilisations with
        $\alpha \geq 1$ yr$^{-1}$, $\tau_d$ falls below $\sim 20$ yr.
  \item The Great Silence may reflect not a scarcity of advanced civilisations but
        the brevity of their detectable phases. 
  \item Current narrowband SETI is already mismatched to our own evolving technology.
        Future searches should be complemened by broadband approaches, persistent
        technology-agnostic signatures, and AI-driven anomaly detection across
        deep, wide-field, electromagnetic/multi-messenger datasets.
\end{enumerate}

The search for extraterrestrial intelligence is, ultimately, a search for our own
future.  As humanity approaches potential post-biological transitions, recognising that
the universe's most advanced civilisations may operate beyond our current observational
horizon is not a counsel of despair, but a call to broaden our strategy.

\section*{Acknowledgments}

\noindent The author would like to thank the symposium organisers for an excellent meeting. 


\end{document}